\documentclass[aps,prb,twocolumn,superscriptaddress]{revtex4-1}
\usepackage{graphics}
\usepackage{graphicx}
\pdfoutput=1
\usepackage{dcolumn}
\usepackage{bm}
\usepackage{float}
\usepackage{stmaryrd}
\usepackage{wasysym}
\usepackage{pifont}

\begin{document}

\title{Properties of single crystalline $A$Zn$_2$Sb$_2$ ($A$=Ca,Eu,Yb)}

\author{Andrew F. May}
\email{mayaf@ornl.gov}
\author{Michael A. McGuire}
\affiliation{Materials Science and Technology Division, Oak Ridge National Laboratory, Oak Ridge, TN 37831}
\author{Jie Ma}
\affiliation{Quantum Condensed Matter Division, Oak Ridge National Laboratory, Oak Ridge, TN 37831}
\author{Olivier Delaire}
\affiliation{Quantum Condensed Matter Division, Oak Ridge National Laboratory, Oak Ridge, TN 37831}
\author{Ashfia Huq}
\affiliation{Quantum Condensed Matter Division, Oak Ridge National Laboratory, Oak Ridge, TN 37831}
\author{Radu Custelcean}
\affiliation{Chemical Sciences Division, Oak Ridge National Laboratory, Oak Ridge, TN 37831}
\date{\today}

\begin{abstract}
Single crystals of CaZn$_2$Sb$_2$, EuZn$_2$Sb$_2$, and YbZn$_2$Sb$_2$ were grown from melts of nominal composition $A$Zn$_5$Sb$_5$ ($A$=Ca,Eu,Yb) with the excess melt being removed at 1073\,K.  The electrical transport properties are consistent with those previously reported for polycrystalline samples.  This confirms that the $p$-type carrier concentrations ranging from 2$\times$10$^{19}$cm$^{-3}$ to $\sim 1\times$10$^{20}$cm$^{-3}$ are intrinsic to these materials.  Also consistent with transport in polycrystalline materials, the carrier mobility is found to be lowest in CaZn$_2$Sb$_2$, suggesting the trends in mobility and thermoelectric efficiency within these compounds are inherent to the material systems and not due to inhomogeneity or impurities in polycrystalline samples.  These results suggest CaZn$_2$Sb$_2$ has the strongest coupling between the doping/defects and the electronic framework.  Magnetization measurements reveal an antiferromagnetic transition near 13\,K in EuZn$_2$Sb$_2$, and the observed magnetic anisotropy indicates the spins align parallel and anti-parallel to $c$ in the trigonal lattice.  Powder neutron diffraction on polycrystalline samples of CaZn$_2$Sb$_2$ and YbZn$_2$Sb$_2$ reveals smooth lattice expansion to 1000\,K, with $c$ expanding faster than $a$.  The Debye temperatures  calculated from specific heat capacity data and the isotropic displacement parameters are found to correlate with the carrier mobility, with the CaZn$_2$Sb$_2$ displaying the largest Debye temperature and smallest mobility.
\end{abstract}

\maketitle

\section{Introduction}

Several antimonides with the CaAl$_2$Si$_2$ structure-type have been shown to possess promising thermoelectric performance.\cite{122Jeff,EuZn2Sb2,Cd122Grin10,Cd122Grin10b,YbCdMnSb2,EuCdZnSb2,122Tober}  The thermoelectric figure of merit $zT=(\alpha^2T)/(\rho\kappa)$ $>$ 1.0 in these materials, with $zT\sim1.2$ reported in YbCd$_{1.6}$Zn$_{0.4}$Sb$_2$ near 700\,K.\cite{Cd122Grin09}  Here, $\alpha$ is the Seebeck coefficient, $\rho$ the electrical resistivity, and $\kappa$ the thermal conductivity.  As this material-specific $zT$ approaches infinity, the efficiency of a corresponding thermoelectric device can theoretically approach the Carnot limit.  While $zT$ is approaching 1.5 or 2 for engineered materials and composites,\cite{SootsmanReview} finding and understanding materials that inherently have $zT$ near unity is still of great interest.\cite{ComplexTE}  Achieving this goal with environmentally friendly elements is of great importance.

As suggested by the formulation of $zT$, large thermoelectric efficiency originates in the coupling of a high electrical conductivity with a large Seebeck coefficient, as well as a low thermal conductivity.  Large Seebeck coefficients are typically associated with insulating behavior due to the inherently large asymmetry in the electrical conductivity about the chemical potential when the chemical potential is located within the energy gap (but not near the center of the gap).  Large electrical conductivity $\sigma=1/\rho=ne\mu$ is generally derived from a large carrier density ($n$) and/or a large carrier mobility ($\mu$).  Optimizing thermoelectric efficiency in a simple semiconductor therefore requires balancing the changes in $\alpha^2$ and $\sigma$ with $n$.  The optimized carrier density and $zT$ depend on the lattice thermal conductivity ($\kappa_L$), the effective band mass ($m^*$), and the carrier mobility.\cite{Ioffe,ComplexTE} 

From a materials point of view, the crystal structure and chemistry determine $m^*$, which can be modified via chemical manipulation.  The carrier mobility is influenced by scattering mechanisms and the band mass, as suggested by the classical expression $\mu=e\tau/m^*$, where $\tau$ is the carrier relaxation time and $e$ the carrier charge.  The influence of $m^*$ is not as clear as it appears, though, as $\tau$ generally depends on $m^*$ as well.\cite{Fistul,WoodReview}  Finally, the values of $\kappa_L$ are also linked to both scattering mechanisms and fundamental material properties.

Understanding the physical mechanisms that influence transport is essential for developing and discovering materials with large thermoelectric efficiency.  One of the best ways to probe these mechanisms is through the characterization of single crystalline materials.  In general, defects and inhomogeneity are minimized in single-crystals, thereby allowing the underlying physics to be revealed.  

Previous studies have shown polycrystalline $A$Zn$_2$Sb$_2$ ($A$=Ca,Sr,Yb,Eu) to display an interesting trend in $\mu$: the  Eu and Yb compounds possess higher carrier mobility than the Ca and Sr compounds.\cite{122Tober}  This results in larger $zT$ in EuZn$_2$Sb$_2$ than in CaZn$_2$Sb$_2$.  These nominally charge-balanced compounds are found to be $p$-type semiconductors with extrinsic carrier concentrations of approximately 2$\times$10$^{19}$cm$^{-3}$ in $A$=Sr,Eu,Ca, and 1$\times$10$^{20}$cm$^{-3}$ for $A$=Yb.  The differences in mobility have been considered from the point of view of the electronic structure, and the band masses appear to be quite similar in these materials.\cite{122Tober}  Therefore, scattering effects are likely responsible for the variations in carrier mobility.  However, it is not known if these trends are intrinsic to the compounds, or are the result of measurements on multiphase, polycrystalline materials.  It is also unclear if the $p$-type carrier concentrations are inherent to the compounds or are due to impurities or grain boundaries.

The current study aims to examine the carrier mobility in single-crystalline $A$Zn$_2$Sb$_2$ for $A$= Ca, Eu, and Yb. Single crystal x-ray diffraction is used to examine crystal quality and investigate the possibility of a structural origin to the mobility trends, and neutron powder diffraction was used to follow the structural properties as a function of temperature.  Heat capacity and magnetization measurements are also utilized to examine the fundamental properties of these materials.

\section{Methods}

Single crystals of CaZn$_2$Sb$_2$, EuZn$_2$Sb$_2$, and YbZn$_2$Sb$_2$ were obtained using the metal-flux method.  High-purity elements were combined in Al$_2$O$_3$ crucibles with nominal compositions $A$Zn$_5$Sb$_5$.  A second crucible filled with quartz wool was utilized to catch the excess molten metal, and this assembly was sealed inside an evacuated quartz ampoule.  The elements were melted together at 1000$^{\circ}$C for 12\,h, and cooled to 800$^{\circ}$C over 96\,h at which point they were removed from the furnace, inverted, and placed into a centrifuge to remove the excess ZnSb.  Similar properties were obtained from crystals of CaZn$_2$Sb$_2$ and YbZn$_2$Sb$_2$ that were kept in the furnace at 800$^{\circ}$C for five days before centrifugation, suggesting the carrier concentration and mobility are inherent to the compounds at these crystallization conditions.

The crystals typically grew with one large, flat-face that was identified as the \textit{ab}-plane through two-circle x-ray diffraction.  This facet was used to orient the crystals, which were ground using fine grit sandpaper and cut by a diamond wafering blade to obtain the geometry necessary for four-point electrical property measurements.  The crystals did not show obvious signs of oxidation over the period of several weeks.

Crystal structures were characterized at 173\,K by single crystal x-ray diffraction using a Bruker SMART APEX CCD diffractometer with Mo-$K\alpha$ radiation ($\lambda$ = 0.71073\,\AA).  Absorption corrections were applied with SADABS and the data were refined using SHELXL-97,\cite{Shelxl97} with ten free parameters for each refinement.  Powder x-ray diffraction data were collected on as-grown facets and hand-ground crystals at ambient conditions on a PANalytical X'Pert Pro MPD using an incident beam Cu K$_{\alpha,1}$ monochromator. 

Neutron powder diffraction measurements were performed on polycrystalline CaZn$_2$Sb$_2$ and YbZn$_2$Sb$_2$ from 10\,K to 973\,K, using the POWGEN time-of-flight diffractometer at the Spallation Neutron Source, Oak Ridge National Laboratory.  Measurements above room temperature occurred under dynamic vacuum in unsealed containers. Subsequent measurements below 300\,K were performed with a partial atmosphere of helium gas in sealed containers, which were mounted on a closed-cycle refrigerator in vacuum. The measurements covered $d$-spacings from $\sim$0.3 to 3.5\,\AA. A vanadium standard was measured to correct the efficiency of the detectors.\cite{Huq2011} Rietveld refinements were carried out using the GSAS software package and the EXPGUI interface \cite{GSAS,EXPGUI} utilizing $d$-spacings from $\sim$0.5 to 3.0\,\AA.  The refinements possessed R$_{wp}$ of 0.0309 and 0.0331, and $\chi^2$ of 8.398 and 10.67 for CaZn$_2$Sb$_2$ and YbZn$_2$Sb$_2$, respectively, at 300\,K.  Some degradation of the samples occurred at high temperatures, likely due to the presence of small amounts of oxygen in the sample environment.  The materials for neutron powder diffraction were synthesized in Al$_2$O$_3$ crucibles with 2\% excess Ca or Yb.  Melting occurred at 1050$^{\circ}$C and was followed by a 24\,h dwell at 800$^{\circ}$C.  The materials were then ground in a He glove box, pressed into pellets, and annealed at 800$^{\circ}$C for 12\,h.  The samples were ground in a He glove box prior to the measurements.  EuZn$_2$Sb$_2$ was not investigated due to strong absorption.  

Electrical resistivity $\rho$ and Hall coefficient $R_{H}$ measurements were utilized to investigate the electrical properties; the Hall carrier density is $n_H$=1/$R_He$.  These measurements were performed in a Quantum Design Physical Property Measurement System (PPMS), and ohmic contacts were made via 0.025\,mm Pt wires connected to the sample with DuPont 4929N silver paste. Hall coefficients were obtained from a fit of the Hall resistance versus magnetic field, with maximum fields of $\pm$6\,Tesla employed. The specific heat capacity was measured in the PPMS below 200\,K using N-grease.  A Quantum Design Magnetic Properties Measurements System (MPMS) was utilized to examine the magnetic behavior between 300 and 1.9\,K.

\section{Results and Discussion}

\subsection{Crystal Structures}

\begin{figure}
	\centering
\includegraphics[width=3in]{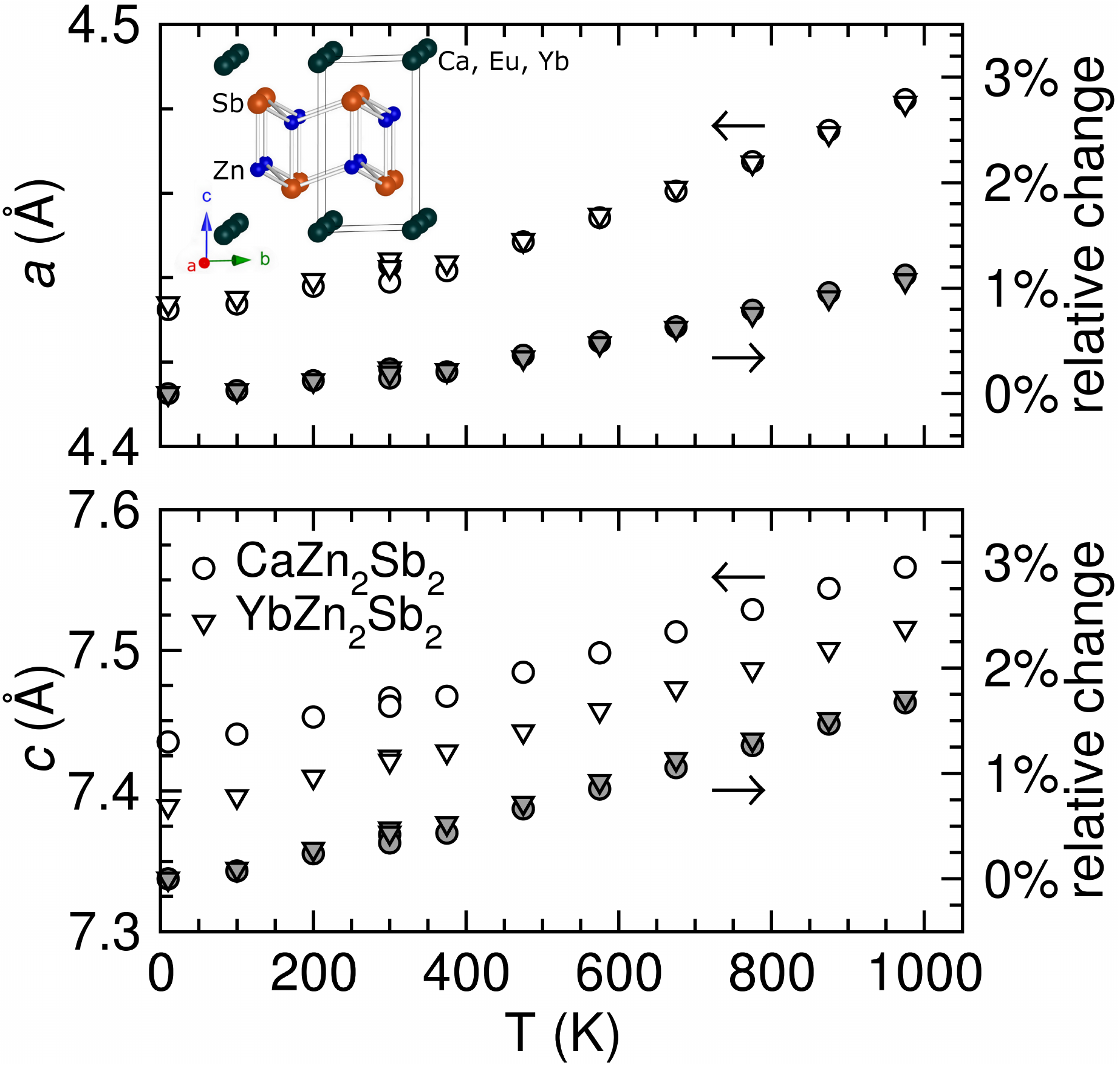}
\caption{Lattice parameters from refinements of powder neutron diffraction data reveal smooth expansion with $c$ expanding more rapidly than $a$.  An inset in the upper panel shows an image of the crystal structure, with unit cell outlined.}
	\label{fig:powgen}
\end{figure}

The $A$Zn$_2$Sb$_2$ compounds are composed of alternating layers of $A^{2+}$ cations and covalent (Zn$_2$Sb$_2$)$^{2-}$ layers, with the layers extending in the \textit{ab}-plane.  An image of the crystal structure is shown as an inset in Figure \ref{fig:powgen}.  The results from refinements of single-crystal x-ray diffraction data agree well with those already presented in the literature,\cite{SrZn2Sb2Structure,MossbauerEu122,122Mar08} and detailed discussions of bonding in this structure type have already been provided.\cite{CaAl2Si2Bonds,BurdettMiller1990}  As shown in Table \ref{tab:refine}, the lattice parameters of EuZn$_2$Sb$_2$ are larger than those of CaZn$_2$Sb$_2$ and YbZn$_2$Sb$_2$, and this trend agrees with the atomic sizes.\cite{PaulingRadii,ShannonRadii}  As the larger Eu atom expands the lattice, the relative change in $c$ is greater than the relative change in $a$.  Similarly, the $A$-Zn interatomic distances change much more than the Zn-Sb bond distances.  Also, the in-plane Sb-$A$-Sb angle changes more than two Sb-Zn-Sb angles.  Similar trends where observed in isostructural $A$Mg$_2$Bi$_2$ crystals ($A$=Ca, Eu, Yb).\cite{AMg2Bi2_Inorg}   The structural changes are biased towards maintaining desired Zn-Sb bond distances and angles.  Contrary to observations in the $A$Mg$_2$Bi$_2$ crystals,\cite{AMg2Bi2_Inorg} however, the Eu-based compound has the least ideal octahedral coordination of Sb about $A$.  In general, the angles quantifying the tetrahedral coordination of Sb about Zn, as well as the octahedral coordination of Sb about $A$, are further from the ideal values for these coordination environments than was observed in $A$Mg$_2$Bi$_2$.  As the same trends were observed for the mobility in both $A$Mg$_2$Bi$_2$ and $A$Zn$_2$Sb$_2$, these small changes in bonding environments do not seem to be strongly correlated with the electrical transport.

The lattice parameters of CaZn$_2$Sb$_2$ and YbZn$_2$Sb$_2$ are found to expand with increasing temperature in a typical manner, as shown in Figure \ref{fig:powgen} for data taken on polycrystalline samples.  The change in $c$ between CaZn$_2$Sb$_2$ and YbZn$_2$Sb$_2$ is consistent with a larger metallic radii for Ca than for Yb.\cite{PaulingRadii}  The expansion in $c$ is larger than expansion in $a$, with relative expansion in $c$ being $\sim$\,49\% and 60\% larger than in $a$ for CaZn$_2$Sb$_2$ and YbZn$_2$Sb$_2$, respectively.  The larger expansion in $c$ is consistent with the layered nature of these compounds, though bonding is fairly three-dimensional and interactions between the $A^{2+}$ and (Zn$_2$Sb$_2$)$^{2-}$ layers are evident in first principles calculations.\cite{122Tober}

\begin{figure}
	\centering
\includegraphics[width=3in]{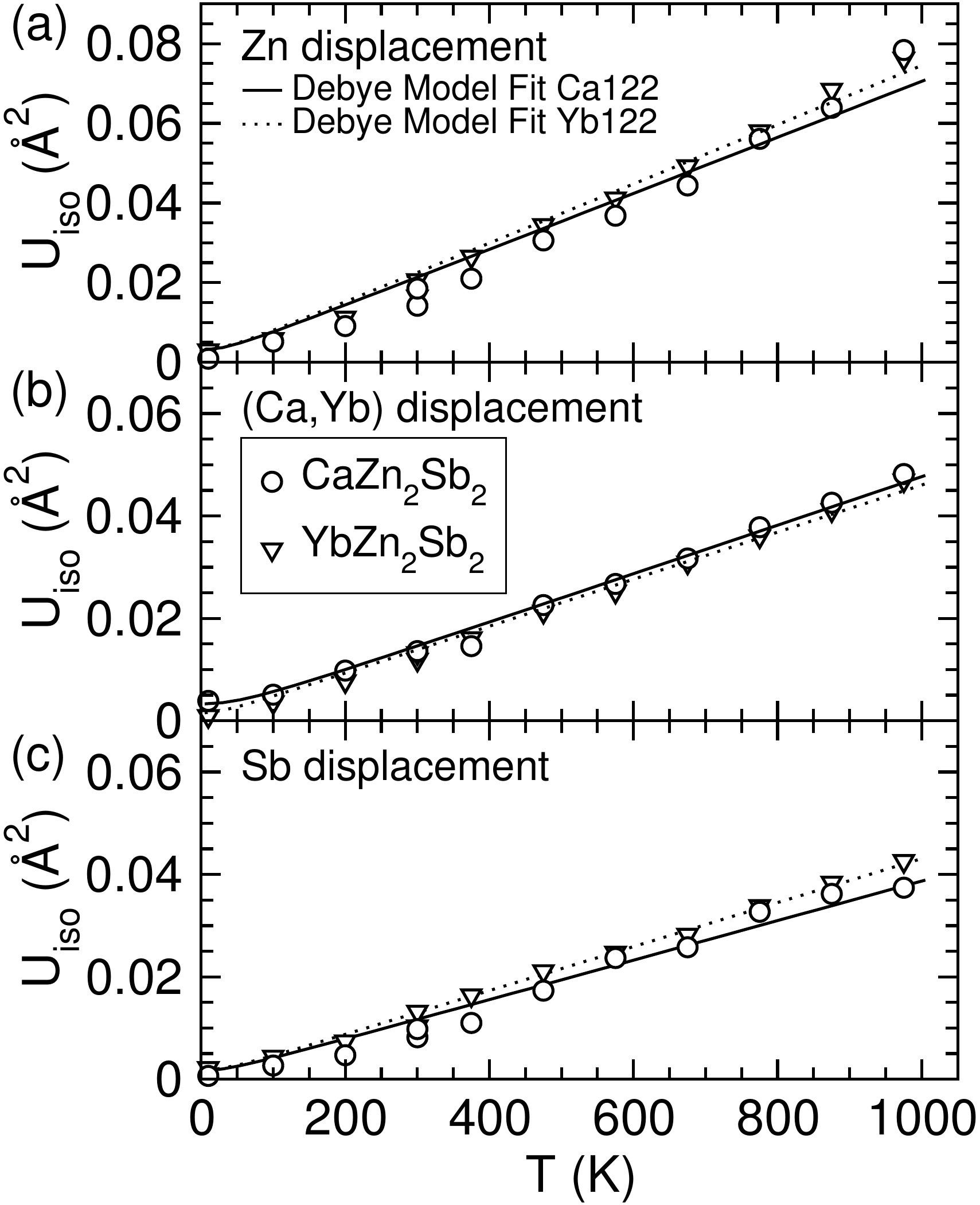}
\caption{Isotropic displacement parameters obtained from refinements of powder neutron diffraction data.  The solid and dashed curves are obtained by fitting the displacement parameters of CaZn$_2$Sb$_2$ and YbZn$_2$Sb$_2$, respectively, using the Debye temperature as a free parameter for each element.}
	\label{fig:U}
\end{figure}

The isotropic displacement parameters (U$_{\textrm{iso}}$) obtained from refinements of neutron powder diffraction data are presented in Figure \ref{fig:U}, where similar values are observed in both compounds.  U$_{\textrm{iso}}$ can be utilized to obtain a Debye temperature ($\Theta_D$), or a theoretical U$_{\textrm{iso}}$ can be obtained from a known $\Theta_D$.  In a monatomic lattice, the Debye model predicts the isotropic displacement parameter to be\cite{WillisPryor}

\begin{equation}
U_{iso} = \frac{3h^2}{4 \pi^2 m \Theta_D}\left(\frac{T^2}{\Theta_D^2}\int_0^{\Theta_D/T}\frac{x}{e^x-1}\textrm{d}x+\frac{1}{4}\right),
\label{eqn:Uiso}
\end{equation}

\noindent where $m$ is the atomic mass. A calculation of U$_{\textrm{iso}}$ using the Debye temperature obtained from specific heat data (below), and the average atomic mass underestimates U$_{\textrm{iso}}$.  Note that the theoretical curves for CaZn$_2$Sb$_2$ and YbZn$_2$Sb$_2$ would be indistinguishable in Figure \ref{fig:U}.   The U$_{\textrm{iso}}$ data can be better described by utilizing a different Debye temperature for each element, as shown by the solid and dotted curves in Figure \ref{fig:U}.  This reveals that the temperature dependences of U$_{\textrm{iso}}$ are well described by the Debye model when $\Theta_D$ is allowed to vary between elements.

The U$_{\textrm{iso}}$ data suggest a larger Debye temperature for CaZn$_2$Sb$_2$ as opposed to YbZn$_2$Sb$_2$, which is consistent with the Debye temperatures obtained from specific heat capacity data.  This difference originates in the displacement of Ca and Yb atoms, which have very similar U$_{\textrm{iso}}$ but atomic masses that differ by more than a factor of four.  The calculated Debye temperature for Ca is 276\,K and 135\,K for Yb, while values between 165\,K and 176\,K are obtained for Zn and Sb in both materials.  The displacement parameter is proportional to $kT/f$, where $f$ is an atomic force constant, and thus the difference in Debye temperatures is associated solely with the change in atomic masses when U$_{\textrm{iso}}$ remains the same. These $\Theta_D$ results come from fitting all of the available data to Equation \ref{eqn:Uiso}, and very similar results are obtained when the high temperature data are fit to the high temperature limit of Equation \ref{eqn:Uiso}, U$_{\textrm{iso}}= \frac{3h^2 T}{4 \pi^2 m k \Theta_D^2}$.  When utilizing average U$_{\textrm{iso}}$ and mean atomic masses, the Debye temperature of the compounds are 180\,K and 153\,K for CaZn$_2$Sb$_2$ and YbZn$_2$Sb$_2$, respectively (high temperature fit).  These values are about 50\,K lower those that obtained from fits of the specific heat capacity, but the trend remains the same.

\begin{table}
  \caption{Selected data from refinements of single crystal x-ray diffraction for CaZn$_2$Sb$_2$, EuZn$_2$Sb$_2$, and YbZn$_2$Sb$_2$ at 173\,K; atomic coordinates of Zn and Sb are at (1/3,2/3,$z$) and Ca, Eu, and Yb are at (0,0,0).}
  \label{tab:refine}
  \begin{tabular}{lccc}
    \hline
   empirical formula & CaZn$_2$Sb$_2$ &  EuZn$_2$Sb$_2$ & YbZn$_2$Sb$_2$\\
   \textit{a}  (\AA)           &  4.4344(4) &  4.4852(4)  &  4.4366(3)   \\
   \textit{c}   (\AA)          &  7.449(1)  &  7.593(1)   &   7.401(1)  \\
   vol (\AA$^3$)               &  126.85(4) & 132.28(3)      &  126.15(3)    \\
   density (g/cm$^3$)          & 5.42       &  6.61       &    7.20       \\
   Zn $z$        							 & 0.63083(7) & 0.63294(9)  &  0.6324(1) \\
   Sb $z$    							     & 0.25701(4) & 0.26533(5)  &   0.25621(6) \\
   U$_{\textrm{eq}}$ Ca/Yb/Eu  & 0.0096(3)  & 0.0090(2)   & 0.0080(2) \\
   U$_{\textrm{eq}}$ Zn        & 0.0106(2)  & 0.0114(2)   & 0.0094(2) \\
   U$_{\textrm{eq}}$ Sb        & 0.0078(2)  & 0.0081(2)   & 0.0061(2) \\
   R$_1$,wR$_2$ (all data)     &  0.0117,0.0259 &   0.0107,0.0281    &    0.0132,0.0292        \\
   reflections,uniq. refl.		 & 881,152    &    893,153  &     877,152    \\
   \hline
  \end{tabular}
\end{table}

\subsection{Electrical Properties}

\begin{figure}
	\centering
\includegraphics[width=3in]{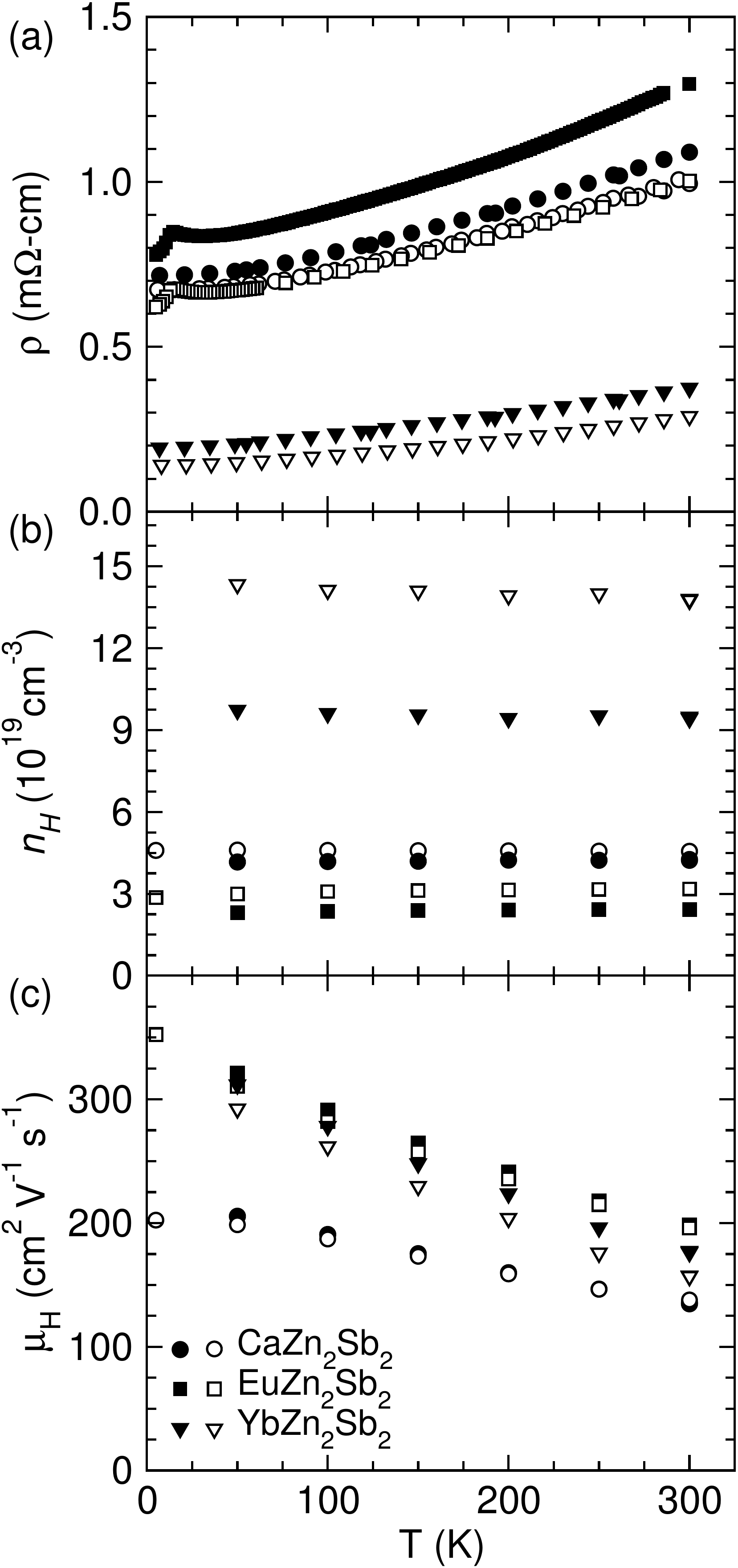}
\caption{Representative temperature dependence of the (a) electrical resistivity, (b) Hall carrier density, and (c) Hall mobility for the $A$Zn$_2$Sb$_2$ single crystals; for each material, data are shown for two crystals, represented by closed and open symbols.}
	\label{fig:elec}
\end{figure}

The electrical transport properties of single crystalline CaZn$_2$Sb$_2$, EuZn$_2$Sb$_2$, and YbZn$_2$Sb$_2$ are shown in Figure \ref{fig:elec}.  Data for two crystals of each composition are included, revealing a distribution of properties observed between various crystals.  Trends between compositions clearly exist, though, as highlighted by the differences in carrier concentrations between the different compositions.

The transport properties of single crystalline CaZn$_2$Sb$_2$, EuZn$_2$Sb$_2$, and YbZn$_2$Sb$_2$ are similar to those reported for polycrystalline samples near room temperature.\cite{122Jeff,122Tober,EuZn2Sb2}  The mobility in EuZn$_2$Sb$_2$ and YbZn$_2$Sb$_2$ are larger than in CaZn$_2$Sb$_2$, and the trend for larger carrier density in YbZn$_2$Sb$_2$ is also consistent with the polycrystalline results (see Table \ref{tab:props}).  The mobility in single-crystalline CaZn$_2$Sb$_2$ and YbZn$_2$Sb$_2$ is larger than that in the polycrystalline CaZn$_2$Sb$_2$ and YbZn$_2$Sb$_2$, suggesting grain boundary scattering or scattering by impurity phases may have reduced the mobility of the polycrystalline samples.  The mobility in single crystals of EuZn$_2$Sb$_2$ is lower than that reported in the polycrystalline materials.  However, the same trends in $\mu$ are observed between compositions for both polycrystalline and single crystalline materials, suggesting the reduced $\mu$ in CaZn$_2$Sb$_2$ is inherent to this compound.

The carrier densities observed here for single crystalline samples are very similar to those reported by various research groups for polycrystalline samples grown from near-stoichiometric melts.  Thus, while the current crystals were grown in excess ZnSb, which may promote the growth of $A$ deficient samples, the carrier densities observed appear to be intrinsic properties of these materials.   Vacancies of either $A$ or Zn would result in holes, and the observation of temperature-independent $n_H$ values are consistent with this hypothesis.  Interestingly, similar trends in $\mu$ and $n_H$ have been found in isostructural $A$Mg$_2$Bi$_2$ single crystals.\cite{AMg2Bi2_Inorg}  Identifying the vacant species is very difficult given the small concentration of vacancies (at most 1/100) required to produce the observed carrier densities. Refinement of single crystal x-ray diffraction data did not provide insight into this problem.  It seems likely, however, that $A$ vacancies are responsible for the free carriers, particularly due to the trends in carrier mobility.

While phonon scattering is clearly the dominant carrier scattering mechanism at high temperatures, the mobility is strongly influenced by a temperature independent scattering mechanism.  This is readily observed via large residual resistances in Figure \ref{fig:elec}a.  As observed in Figure \ref{fig:elec}c, this scattering mechanism results in suppressed $\mu_H$ for CaZn$_2$Sb$_2$ for all temperatures examined, and thus results in reduced $zT$ as well.  In analogous CaMg$_2$Bi$_2$ crystals, the mobility decreases at low $T$, suggesting a strong temperature-dependent scattering mechanism.\cite{AMg2Bi2_Inorg}  The influence of the defect scattering can be quantified by the resistivity ratios.  The magnetic transition in EuZn$_2$Sb$_2$ makes this more difficult, thus consider $\rho$(300\,K)/$\rho$(20\,K) for simplicity.  This ratio is smallest for CaZn$_2$Sb$_2$ and EuZn$_2$Sb$_2$ ($\rho$(300\,K)/$\rho$(20\,K) $\approx$ 1.5) while it is larger for YbZn$_2$Sb$_2$ ($\rho$(300\,K)/$\rho$(20\,K) $\approx$ 1.9).  Interestingly, this scattering mechanism is less detrimental to the mobility in YbZn$_2$Sb$_2$ despite the inferred increase in defect concentration (larger carrier concentration). 

The only material parameter that correlates with the mobility is the Debye temperature; larger mobility is observed in conjunction with lower Debye temperatures.  As discussed above, the Debye temperature can be related to the thermal displacement parameters in a solid and can therefore provide insight into local bonding environments.  Classically, the similar displacement parameters equate to similar force constants for Ca and Yb, though the mass difference points to a larger maximum velocity of vibration for Ca.  However, the displacement parameters are not unusually larger and it appears that defects are responsible for the variations in mobility.  The trends in $\Theta_D$ provide insight into the local bonding environments and may help understand how defects influence the surrounding lattice and the corresponding electronic structure/transport.  Perhaps this system will be viewed by theorists as an approachable and rewarding challenge.

The transition from the paramagnetic to antiferromagnetic (AFM) state in EuZn$_2$Sb$_2$ is observed in the electrical resistivity data.  The transition occurs near 13\,K, which is consistent with previous observations.\cite{EuZn2Sb2,PhysRevB.73.014427}  The transition causes a small increase in $\rho$ at temperatures just above the transition, and a sharp decrease in $\rho$ just below the transition temperature.  This behavior  is associated with increased spin-disorder scattering near the transition, and a sharp reduction in this scattering below the transition.

\begin{table}
\caption{Room temperature transport data for single crystalline samples compared to the literature for polycrystalline samples of CaZn$_2$Sb$_2$ and YbZn$_2$Sb$_2$,\cite{122Jeff,Cd122Grin09} and EuZn$_2$Sb$_2$\cite{EuZn2Sb2}.\label{tab:props}}
\begin{tabular}[c]{|c|c|c|c|c|}
\hline
 Compound & poly/single &  $n_H$ (300\,K) & $\mu_H$ (300\,K) & $\rho$ (300\,K) \\
 - & - & 10$^{19}$cm$^{-3}$ & cm$^{2}$V$^{-1}$s$^{-1}$  & m$\Omega$ cm  \\
\hline
 CaZn$_2$Sb$_2$ & single &   4.2--4.6  & 135       & 1.0--1.1 \\
 CaZn$_2$Sb$_2$ & poly   &    3.1      & 83        & 2.4  \\
 EuZn$_2$Sb$_2$ & single &    2.4--3.2 & 200       & 1.0--1.3 \\
 EuZn$_2$Sb$_2$ & poly   &    2.8      & 257       & 0.88  \\
 YbZn$_2$Sb$_2$ & single &   9.4--14   & 157--175  & 0.29--0.37 \\
 YbZn$_2$Sb$_2$ & poly   &   11.4--15  & 120--130  & 0.3-0.4 \\
\hline
\end{tabular}
\end{table}

\subsection{Heat Capacity}

\begin{figure}
	\centering
\includegraphics[width=3in]{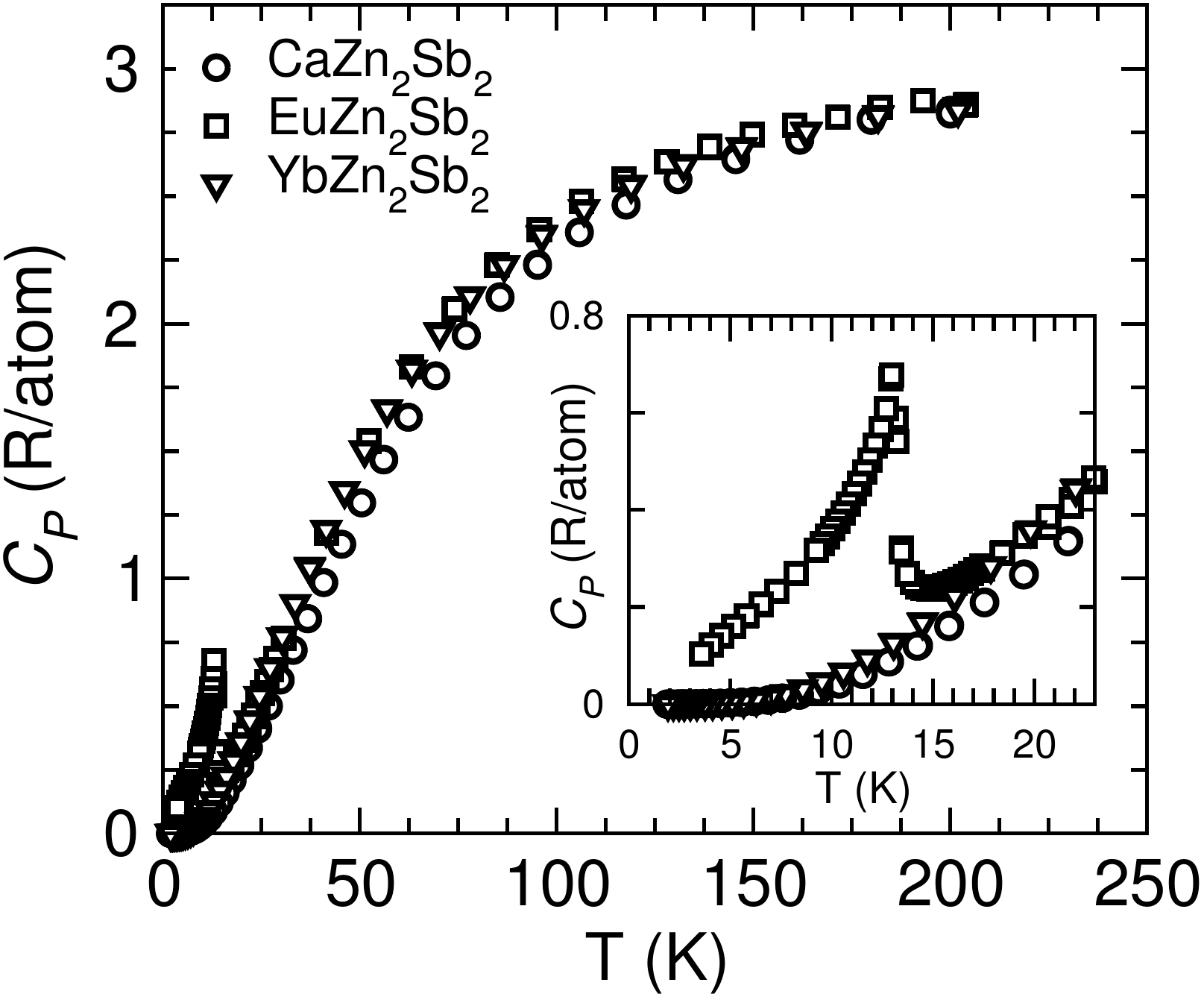}
\caption{Specific heat capacity for CaZn$_2$Sb$_2$, EuZn$_2$Sb$_2$, and YbZn$_2$Sb$_2$ with inset highlighting the transition in EuZn$_2$Sb$_2$ $\sim$ 13\,K.}
	\label{fig:Cp}
\end{figure}

Experimental values of the specific heat capacity ($C_P$) are shown in Figure \ref{fig:Cp}, and $C_P$ is approaching the simple theoretical limit of 3R/atom for all compounds.  The antiferromagnetic transition in EuZn$_2$Sb$_2$ is clearly shown in the inset.  By utilizing the YbZn$_2$Sb$_2$ data as a baseline, the entropy change associated with this magnetic transition is estimated via $\Delta S = \int_{4}^{20} \frac{C_{P,Eu122}-C_{P,Yb122}}{T}\mathrm{d}T \approx$ 13\,J/mol/K.  This value is similar in magnitude to that expected for a transition from the high-entropy paramagnetic phase to the low entropy AFM phase.  The entropy change for this type of phase transition is large for compounds containing divalent Eu, which has $J = \frac{7}{2}$; simple theory estimates the entropy of a paramagnet as $S = R$ln[2J+1],\cite{Smart} which is 17.3\,J/mol/K for  $J = \frac{7}{2}$.

The heat capacity of CaZn$_2$Sb$_2$ is observed to be lower than that of EuZn$_2$Sb$_2$ and YbZn$_2$Sb$_2$, and this trend is present down to low temperatures (inset Figure \ref{fig:Cp}).  Similar trends were also observed in single crystals of CaMg$_2$Bi$_2$, EuMg$_2$Bi$_2$ and YbMg$_2$Bi$_2$.\cite{AMg2Bi2_Inorg}  The suppression of $C_P$ can be regarded as an increase in the Debye temperature $\Theta_D$.  For a solid with $n$ atoms per formula unit, the Debye model expresses the heat capacity in terms of the reduced phonon energy $x$ as

\begin{equation}
C = 9 R n \left(\frac{T}{\Theta_D}\right)^3 \int^{\Theta_D/T}{\frac{x^4 e^x}{(e^x-1)^2}\mathrm{d}x}.
\label{eqn:Cp}
\end{equation}

\noindent While the Debye model fails to quantitatively describe the data over a wide temperature range, it is particularly useful when comparing similar compounds.  By fitting the heat capacity above 20\,K, we find that $\Theta_D$ for CaZn$_2$Sb$_2$ is 224\,K, while for EuZn$_2$Sb$_2$ is it 195\,K and for YbZn$_2$Sb$_2$ it is 201\,K.  The Debye temperature obtained by fitting the low temperature data is 234\,K for CaZn$_2$Sb$_2$ and 204 for YbZn$_2$Sb$_2$.  The low temperature analysis utilizes $C_P$= $\gamma T$ + $\beta T^3$, where $\beta$= 12 $\pi^4 R N_{at} /(5 \Theta_D^3$) and $\gamma$ is the Sommerfeld coefficient describing the electronic contribution, which is small in these materials.

The Debye temperature can be related to the material properties via

\begin{equation}
\Theta_D = \frac{h}{k}\left(\frac{3 n N_{a} d}{4 \pi M}\right)^{\frac{1}{3}}v_m,
\label{eqn:Theta}
\end{equation}

\noindent where $N_a$ is Avogadro's number, $d$ the density, $v_m$ the mean speed of sound, and $M$ the molecular weight.\cite{DebyeTempAnderson}  While the value of $M$ changes by a fairly large percentage between these compounds, the ratio $d/M$ changes by less than 5\% between compounds and thus the changes in atomic mass alone appear to be insufficient to account for the changes in Debye temperatures.  Therefore, the variations in $\Theta_D$ likely represent variations in $v_m$, suggesting the phonon velocities in CaZn$_2$Sb$_2$ are larger than those in EuZn$_2$Sb$_2$ and YbZn$_2$Sb$_2$ by roughly 15\%.  Measurements in polycrystalline materials do suggest the lattice thermal conductivity ($\kappa_L$) is larger in CaZn$_2$Sb$_2$ than in EuZn$_2$Sb$_2$, which is consistent with this trend in $\Theta_D$.  Note that the major difference in thermoelectric efficiency between CaZn$_2$Sb$_2$ and EuZn$_2$Sb$_2$ is associated with the larger mobility in EuZn$_2$Sb$_2$, and not the thermal conductivity.

\subsection{Magnetic Properties}

\begin{figure}
	\centering
\includegraphics[width=3in]{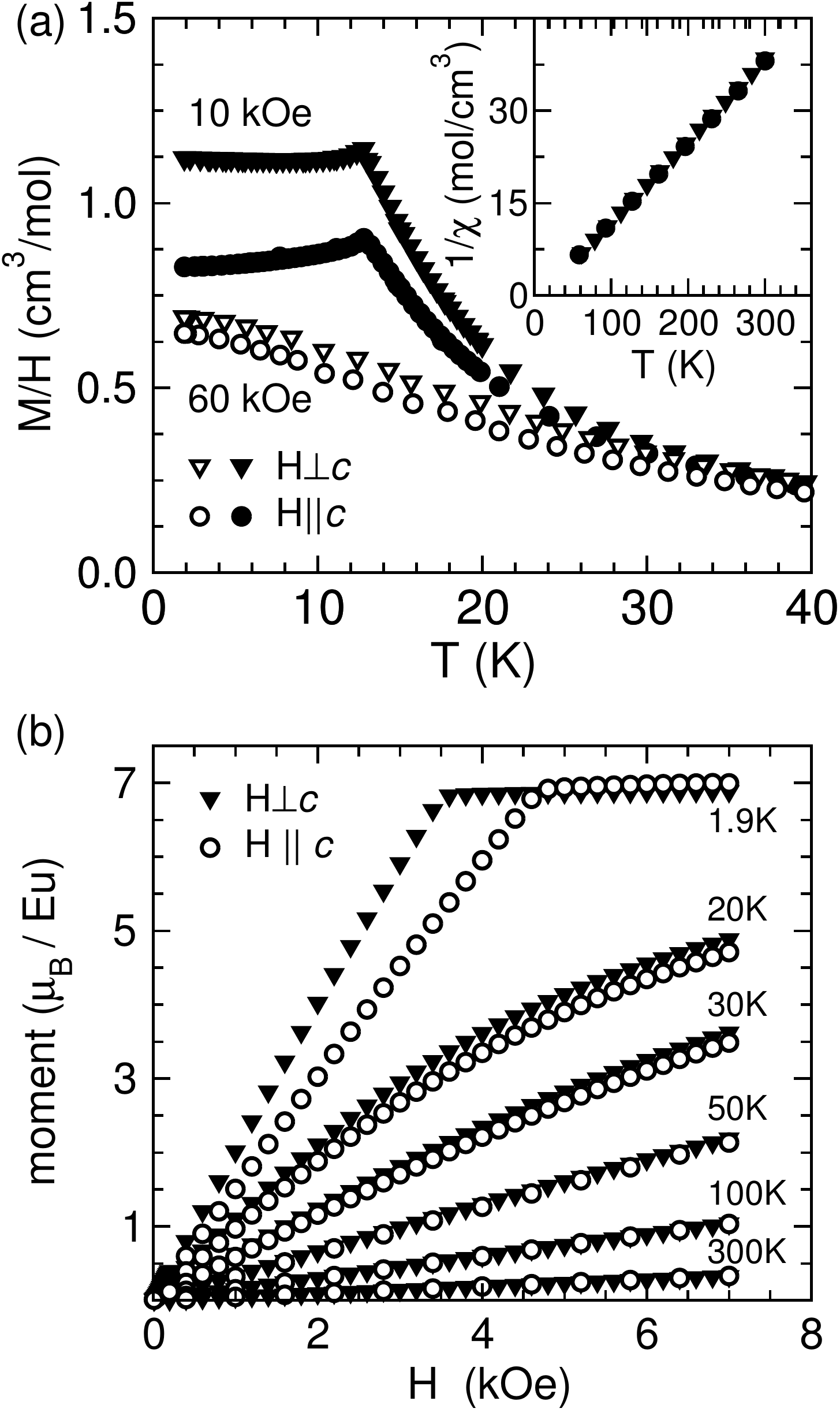}
\caption{(a) The antiferromagnetic transition near 13\,K in EuZn$_2$Sb$_2$ is observed via the temperature dependence of \textbf{M/H} in moderate fields; the transition is suppressed at high fields.  (b) Field dependence of the induced moment showing a saturation near the expected 7\,$\mu_B$/Eu below the AFM transition, while linear behavior consistent with paramagnetism is observed at higher temperatures.}
	\label{fig:mag}
\end{figure}

The magnetization data for EuZn$_2$Sb$_2$ are shown in Figure \ref{fig:mag}, and the data are consistent with literature reports on polycrystalline samples, as well as on a matrix of ZnSb/EuZn$_2$Sb$_2$.\cite{EuZn2Sb2,PhysRevB.73.014427,MossbauerEu122}  A cusp in the temperature-dependence of the moment \textbf{M/H} is observed for low to moderate fields near 13\,K, which is associated with a transition from paramagnetism to an antiferromagnetic ordering of the Eu moments.  The sharp feature is suppressed at higher fields, as shown in Figure \ref{fig:mag}a for measurements in 60\,kOe.

At \textbf{H}=0, the Eu moments order antiferromagnetically with spins parallel (and anti-parallel) with the $c$-axis (below $\sim$13\,K).  Application of a field continuously reorients the moments until they are aligned with the external field, and the induced moment saturates near the expected 7$\mu_B$/Eu (Figure \ref{fig:mag}b).  The orientation of the spins is revealed by the near temperature-independent values of \textbf{M/H} below 13\,K for \textbf{H} $\perp$ $c$, as well as the decrease in \textbf{M/H} with decreasing $T$ for \textbf{H} $\parallel$ $c$ below 13\,K (at 10\,kOe).  A larger critical field is observed for \textbf{H} $\parallel$ $c$ than for \textbf{H} $\perp$ $c$, and the associated reduction in \textbf{M/H} is observed in Figure \ref{fig:mag}a for \textbf{H}$\parallel$$c$.

%\begin{figure}
%	\centering
%\includegraphics[width=3in]{Mag2PanelZnSb.pdf}
%\caption{(a) The antiferromagnetic transition near 13\,K in EuZn$_2$Sb$_2$ is observed via the temperature dependence of \textbf{M/H} in moderate fields; the transition is suppressed at high fields.  (b) Field dependence of the induced moment showing a saturation near the expected 7\,$\mu_B$/Eu below the AFM transition, while linear behavior consistent with paramagnetism is observed at higher temperatures.}
%	\label{fig:mag}
%\end{figure}

At temperatures well above the transition, isotropic paramagnetic behavior is observed (inset Fig. \ref{fig:mag}a, and linear behavior in Fig. \ref{fig:mag}b at high $T$).  In this region, the susceptibility data are described by a modified Curie-Weiss law, with analysis between 50 and 300\,K yielding Weiss temperatures of approximately -7\,K and an effective moment of 7.93(1)$\mu_B$/Eu.  This is very close to the theoretical effective moment of 7.94$\mu_B$/Eu$^{2+}$.

Magnetization measurements on YbZn$_2$Sb$_2$ revealed a weak diamagnetic signal consistent with divalent Yb.  At low temperatures, a small paramagnetic component was observed, possibly due to magnetic impurities or the existence of some trivalent Yb.  Analyzing the low-temperature data under the assumption that all of the paramagnetic component is due to Yb$^{3+}$ yields an upper bound of $\sim1\%$ Yb$^{3+}$.  It is difficult to rule out trivalent Yb in these compounds, particularly because techniques aimed at identifying the valence of Yb are strongly influenced by oxidation.\cite{XPS_Yb122_Espen}

\subsection{Summary}

Single crystals of CaZn$_2$Sb$_2$, EuZn$_2$Sb$_2$, and YbZn$_2$Sb$_2$ are found to have electrical properties similar to those reported for polycrystalline samples.  Therefore, the trend for larger carrier mobility in the rare-earth derived compounds appears to be an inherent feature of these materials, and is not related to sample purity.  This confirms that larger $zT$ in EuZn$_2$Sb$_2$ is due to inherent material properties.  Furthermore, the carrier densities observed are similar to those reported for polycrystalline samples, suggesting the dominant defects in these materials are intrinsic to the phases studied.  These defects strongly influence the carrier mobility, particularly at low temperatures, and understanding their existence and influence on transport is necessary if large enhancements in the thermoelectric efficiency of these materials are to be achieved.  This study also reveals that the compounds with larger mobilities have lower Debye temperatures, and further investigations of this correlation are warranted.

\section{Acknowledgements}

This work was supported by the U. S. Department of Energy, Office of Basic Energy Sciences, Materials Sciences and Engineering Division (A.F.M., M.A.M.).  R.C. was supported by the U.S. Department of Energy, Office of Basic Energy Sciences, Division of Chemical Sciences, Geosciences, and Biosciences. The research at Oak Ridge National Laboratory's Spallation Neutron Source was sponsored by the Scientific User Facilities Division, Office of Basic Energy Sciences, U.S. Department of Energy.  O.D. and J.M. were supported by the U. S. Department of Energy, Office of Basic Energy Sciences, through the S3TEC Energy Frontier Research Center, Department of Energy DESC0001299.

\end{document}